\documentclass[manuscript]{aastex}

\shorttitle{O K-edge absorption}
\shortauthors{Gatuzz et al.}

\begin{document}

\title{Physical properties of the interstellar medium using high-resolution Chandra spectra: O K-edge absorption}

\author{E. Gatuzz\altaffilmark{1},
        J. Garc\'ia\altaffilmark{2},
        C. Mendoza\altaffilmark{1,3},
        T.~R. Kallman\altaffilmark{4},
        M.~A. Bautista\altaffilmark{3},
        T.~W. Gorczyca\altaffilmark{3}
        }

\altaffiltext{1}{Centro de F\'isica, Instituto Venezolano de Investigaciones Cient\'ificas (IVIC), PO Box 20632, Caracas 1020A,  Venezuela
\email{egatuzz@ivic.gob.ve, claudio@ivic.gob.ve}}

\altaffiltext{2}{Harvard-Smithsonian Center for Astrophysics, MS-6, 60 Garden Street, Cambridge, MA 02138, USA
\email{javier@head.cfa.harvard.edu}}

\altaffiltext{3}{Department of Physics, Western Michigan University, Kalamazoo, MI 49008, USA
\email{manuel.bautista@wmich.edu, thomas.gorczyca@wmich.edu}}

\altaffiltext{4}{NASA Goddard Space Flight Center, Greenbelt, MD 20771, USA
\email{timothy.r.kallman@nasa.gov}}

\begin{abstract}
  {\em Chandra} high-resolution spectra toward eight low-mass Galactic binaries have been analyzed with a photoionization model that is capable of determining the physical state of the interstellar medium. Particular attention is given to the accuracy of the atomic data. Hydrogen column densities are derived with a broadband fit that takes into account pileup effects, and in general are in good agreement with previous results. The dominant features in the oxygen-edge region are \ion{O}{1} and \ion{O}{2} K$\alpha$ absorption lines whose simultaneous fits lead to average values of the ionization parameter of $\log\xi=-2.90$ and oxygen abundance of $A_{\rm O}=0.70$. The latter is given relative to the standard by \citet{gre98}, but rescaling with the revision by \citet{asp09} would lead to an average abundance value fairly close to solar. The low average oxygen column density ($N_{\rm O}=9.2 \times 10^{17}$~cm$^{-2}$) suggests a correlation with the low ionization parameters, the latter also being in evidence in the column density ratios $N$(\ion{O}{2})/$N$(\ion{O}{1}) and $N$(\ion{O}{3})/$N$(\ion{O}{1}) that are estimated to be less than 0.1. We do not find conclusive evidence for absorption by any other compound but atomic oxygen in our oxygen-edge region analysis.
\end{abstract}


\keywords{atomic processes --- ISM: abundances --- ISM: atoms --- X-rays: binaries --- X-rays: ISM}


\section{Introduction}\label{sec_intro}

High-resolution X-ray spectroscopy provides a powerful technique for studying the interstellar medium (ISM) since the analysis of absorption features allows the identification of atomic and molecular transitions directly related to its physical and chemical properties. In this context, a chemical element of much interest is oxygen due to its cosmic abundance and foremost spectral features, namely K-edge structures whose models can lead to reliable estimates of its abundance, column densities, and ionization fractions.

A prominent oxygen K edge and hints of a resonant $1s\rightarrow 2p$ absorption line were reported in the pioneering work by \citet{sch86} on the Crab Nebula soft X-ray spectrum ($<3$~keV) taken with the Einstein Observatory. No apparent evidence for depletion was therein found thus implying dust grain sizes of $< 4\,\mu$m. A further attempt to exploit the oxygen K-shell absorption region was performed by \citet{jue04}, who analyzed the spectra of seven X-ray binaries taken with the satellite-borne {\it Chandra} High Energy Transmission Grating Spectrometer (HETGS). Their model included two absorption edges ($1s2s^22p^4\ ^4P,\ ^2P)$ and five Gaussians corresponding to K$\alpha$ ($1s\rightarrow 2p$) transitions in \ion{O}{1}, \ion{O}{2}, and \ion{O}{3} and K$\beta$ ($1s\rightarrow 3p$) in \ion{O}{1}. As a result, they managed to constrain the oxygen ionization fractions to $N({\rm O\ II})/N({\rm O\ I})\approx 0.1$ and $N({\rm O\ III})/N({\rm O\ I})\leq 0.1$. \citet{yao05} performed a similar analysis including observations of seven Galactic low-mass X-ray binaries taken with both the High Energy Transmission Grating (HETG) and Low Energy Transmission Grating (LETG) on board {\it Chandra}. The \ion{O}{7} and \ion{O}{8} K$\alpha$ lines and the \ion{O}{7} K$\beta$ detected in three of these sources were associated to the hot interstellar medium (HISM) whose distribution was also estimated; however, due to the few lines of sight, such inference could be biased. \citet{cos05} studied high-resolution spectra of the Cygnus~X-2 low-mass X-ray binary taken with the {\it XMM-Newton} space telescope, concluding that the complexity of the oxygen edge could be explained by absorption of both molecular and ionized atomic species. Moreover, in an analysis of the oxygen K-shell edge in {\it XMM-Newton} spectra towards galaxy clusters, \citet{bau06} determined a high gas-to-dust ratio in the Galaxy, pointing out that the composition of denser clouds could be similar to the diffuse ISM.

\citet{mil09b} have fitted {\it Chandra} HETG data from five X-ray binaries with an absorption model referred to as {\tt TBnew}, and suggested that ISM absorption dominated the neutral column density. From similar observations of the line of sight of Cygnus~X-2, \citet{yao09} detected K absorption lines from \ion{O}{1}, \ion{O}{2}, \ion{O}{6}, \ion{O}{7}, and \ion{O}{8}, determining abundances that indicated mild oxygen depletion onto dust grains in the cold phase. From {\it XMM-Newton} Reflection Grating Spectrometer (RGS) spectra of the X-ray binary Sco~X-1, \citet{dev09} have deduced the presence of extended X-ray absorption fine structures (EXAF) and that around $30{-}50\%$ of the oxygen was bound in solid material. However, \citet{gar11} showed that, by using an accurate photoionization cross section for atomic \ion{O}{1}, observations of this source in the $12.5{-}21.0$~\AA\ region could be adequately reproduced with no evidence of additional molecular and dust contributions. \citet{pin10} have considered both the dust and gas components in the ISM absorption features towards the low-mass X-ray binary GS~1826-238; they found oxygen to be over abundant and a gas ionization degree above $5\%$, concluding that at least $10\%$ of the oxygen was found in molecules or dust.

More recently, \citet{pin13} have measured oxygen column densities and estimated abundance gradients from {\it XMM-Newton} grating spectra of nine low-mass X-ray binaries, concluding that $15{-}25\%$ of the oxygen is found in dust. They also point out that the ratios between different ionization stages are similar among the lines of sight suggesting large-scale chemical homogeneity. \citet{lia13} have analyzed 36 {\it Chandra} HETG observations from eleven low-mass X-ray binaries taking into consideration a correction for the Galactic rotation velocity relative to the rest frame. Through a Bayesian statistical approach and a combined fit, they detect several spectral features in the oxygen-edge region including K$\alpha$ lines from \ion{O}{1}, \ion{O}{2}, \ion{O}{3}, \ion{O}{4}, \ion{O}{5}, \ion{O}{6}, and \ion{O}{7}.

\citet{gat13a} have presented a self-consistent physical model of ISM oxygen K absorption to fit four {\it Chandra} spectra of XTE~J1817-330, a low-mass X-ray binary with high signal-to-noise ratio. The oxygen K-edge region is fitted with a photoionization model derived with the {\sc xstar} computer code \citep{bau01} referred to as {\tt warmabs}. It is found that the K lines and photoionization cross sections must be slightly wavelength shifted in order to match the observed spectra. The resulting low ionization parameter indicate a dominant neutral component with absorption lines from \ion{O}{1} K$\alpha$, K$\beta$, and K$\gamma$ together with K$\alpha$ from \ion{O}{2} and \ion{O}{3}. K$\alpha$ lines from \ion{O}{6} and \ion{O}{7} are also detected, but it is argued that they probably arise in the neighborhood of the source rather than in the ISM;  unsurprisingly, molecular features in the O K-edge region are not found in this study.

In the present work we refine the building blocks of our model of ISM oxygen photoabsorption, and explore its possibilities in describing the high-resolution {\it Chandra} spectra of eight X-ray binaries. Since this process is dominated by the neutral, we evaluate the impact of the atomic data on model prediction by considering both the new and definitive \ion{O}{1} cross section by \citet{gor13} and previous data of \citet{gar05}. Although dramatic revelations are not expected with the present line-of-sight sampling, we are at least certain to contribute to the current discussion on ISM phase composition and local variability. The outline of the paper is as follows. In Section~\ref{sec_atom} we give details concerning the atomic data sets, and the observational reduction steps are described in Section~\ref{sec_obs}. The fitting procedures of \citet{gat13a} for the broadband and oxygen-edge regions are recapitulated in Section~\ref{sec_fit} followed by the fit results (Sections~\ref{sec_fit_broad}--\ref{sec_fit_oxygen}). A discussion of results is carried out in Section~\ref{sec_ism} in order to draw some conclusions in Section~\ref{sec_conclusion}.

\section{Atomic data}\label{sec_atom}

A detailed and quantitative description of inner-shell photoabsorption processes is of vital importance to sustain  astrophysical inferences from ISM X-ray absorption spectra, and as recently discussed for oxygen \citep{gor13}, unsatisfactory discrepancies in both experimental and theoretical atomic data have led to a variety of disjoint conclusions regarding ionization and atomic-to-molecular fractions. In this respect, the spectral edge resonance structures, which give rise to useful astrophysical imprints, are particularly difficult to master as they strongly depend on ionization fractions and a variety of subtle relaxation effects such as Auger and radiative damping.

The latter atomic processes were taken into account in the photoabsorption cross sections of oxygen ions computed by \citet{gar05}, which have been incorporated in the {\tt warmabs} physical model (see Section~\ref{sec_fit}). However, their previous application in X-ray binary spectral modeling, namely {\em XMM-Newton} and {\em Chandra} observations of Sco X-1 \citep{gar11} and XTE J1817−330 \citep{gat13a}, required small wavelength shifts of the theoretical lines in order to secure accurate fits. \citet{gor13} have recently made a serious attempt to establish a definitive photoabsorption model for \ion{O}{1} by taking care of all the relevant features---shoulders, line positions, widths, and oscillator strengths---and by providing a handy analytical expression for prospective users. However, a bothersome discrepancy of 0.6~eV still persists between the astronomically observed K-line position and laboratory measurements. Unconventionally and as fully discussed in \citet{gor13}, the final energy scale was calibrated by aligning to the former for the following reasons. First, the astronomical calibration is consistent with that of the well-understood \ion{O}{7} K$\alpha$ line. Second, the accuracy of the molecular data used in the experimental calibration is not altogether clear \citep{sto97,mcl13}, and preliminary measurements appear to indicate that the experimental \ion{O}{1} K-edge calibration may indeed need a revision (W.~C. Stolte, private communications, 2014). Moreover, \citet{gor13} also found a $\sim 10{-}15$\% dispersion in the background oscillator strengths that is now eliminated by benchmarking to the above-threshold experimental results \citep{hen93}. We are of the opinion that the small {\tt warmabs} uncertainties due to these two cross sections finally closes the argument on atomic oxygen data translation.

\section{Observations and data reduction}\label{sec_obs}

In order to study the ISM oxygen edge along different lines of sight, we analyze {\it Chandra} spectra of eight low-mass X-ray binaries. Table \ref{tab1} tabulates the observational specifications, namely IDs, dates, exposure times, and read modes. All spectra have been obtained with the Medium Energy Gratings (MEG) from the HETGS in combination with the Advanced CCD Imaging Spectrometer (ACIS). Observations were taken in continuous clocking mode (CC) or time exposure mode (TE), in the former the temporal resolution is increased in order to minimize the pileup effect \citep{cac08b}.  Pileup is concerned with the detection of two or more photons as a single event, and can produce a deformation of the continuum level and shape and, consequently, must be reduced as much as possible \citep{mil06a}. We have estimated the pileup contribution along the broadband region with the {\tt simplegpile2} procedure. In TE mode, the ACIS instrument reads the collected photons periodically.  In both cases, we have reduced the data using the standard CIAO threads\footnote{http://cxc.harvard.edu/ciao/threads/gspec.html} taking into account that, if the zero-order data were not telemetered, the zero-order position was estimated using the {\tt findzo} algorithm\footnote{http://space.mit.edu/cxc/analysis/findzo/}. Finally, we have used the {\sc isis} package (version 1.6.2-27\footnote{http://space.mit.edu/cxc/isis/}) for spectral fitting.

\section{Fitting procedure}\label{sec_fit}

Following \citet{gat13a}, we estimate the hydrogen column density using an X-ray absorption model combined with a pileup convolution model (hereafter referred to as the broadband fit). We then reproduce the oxygen-edge region using the {\tt warmabs} model to obtain physical properties related to the gas ionization and elemental abundances. In the broadband region (11--24~\AA), we fit all observations for each source simultaneously with the {\tt simplegpile2(TBnew(powerlaw))} model in order to obtain, using $\chi^{2}$ statistics, an estimate of the hydrogen column density. The {\tt simplegpile2} is a convolution model that takes into account pileup effects, and {\tt TBnew} is an X-ray absorption model that includes chemical elements with atomic numbers $1\leq Z\leq 28$ \citep[see][for details]{gat13a}. In addition to the H column density, $N_{\mathrm H}$, we treat the abundances of O, Ne, and Fe as free parameters. We group each spectrum to obtain a 0.1~\AA\ bin size and add Gaussians to fit the remaining absorption lines, if any. We fix the absorption parameters in all the observations and vary the power law, pileup, and Gaussian parameters in each case. We then introduce the photoabsorption cross sections by \citet{ver96} and chemical abundances by \citet{wil00} although the resulting abundances are given relative to \citet{gre98} for consistency with the {\tt warmabs} output.

In the oxygen-edge region (21--24~\AA), we rebin the data to obtain 20 counts per channel in order to proceed with the analysis using $\chi^{2}$ statistics \citep{nou89}. The free variables are the ionization parameter (log $\xi$), oxygen abundance ($A_{\rm O}$), turbulent broadening (in km\,s$^{-1}$), and redshift, while the hydrogen column density is held fixed at the value obtained in the previous broadband fit. For each source, we fit all observations simultaneously using the same {\tt warmabs} data but varying the power-law parameters for each observation. We perform two separate fits: (1) using the photoabsorption cross sections by \citet{gar05} for O ions and (2) using the new cross section by \citet{gor13} for \ion{O}{1} and those by   \citet{gar05} for the other ionized species. As previously mentioned, we intend to assess these atomic cross sections by delimiting the variations they produce in the ISM oxygen abundances and ionization fractions. Abundances are again given relative to the solar standard by \citet{gre98}.

\section{Broadband fit}\label{sec_fit_broad}

\begin{figure}
  \epsscale{0.9}
  \plotone{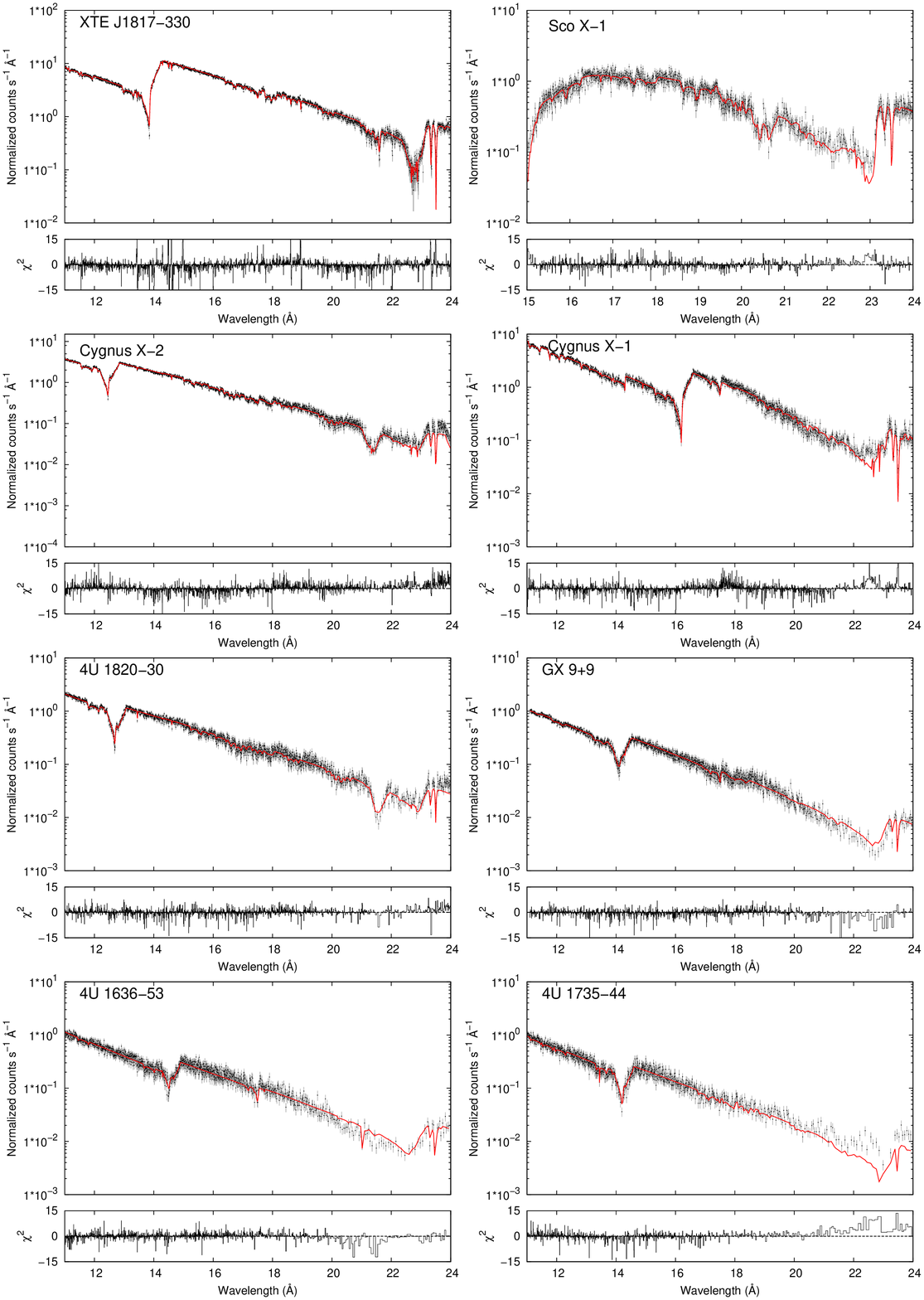}
  \caption{Best broadband fits with the {\tt TBnew} model of the spectra towards eight Galactic low-mass X-ray binaries. For each source the observation with highest signal-to-noise ratio is shown.\label{fig1}}
\end{figure}

Figure~\ref{fig1} shows the broadband fit using the {\tt simple\_gpile2.sl}({\tt TBnew}({\tt powerlaw})) model for each source in the 11--24~\AA\ spectral region. In the case of Sco~X-1 the fit was performed in the  15--24~\AA\ interval as there are no counts below ${\sim}15$~\AA. For each source the observation with the highest signal-to-noise is plotted and residuals in units of $\chi ^2$ are also given. In all cases the Ne-edge absorption region (13--15~\AA) dominates the broadband fit due to a higher photon count (a factor of ${\sim}10$ higher than the photon count in the oxygen edge).

Fit results are given in Table \ref{tab2} including the set for XTE~J1817-330 obtained by \citet{gat13a} where it may be seen that the reduced chi-square values denote a good fit in each case. Oxygen, neon, and iron abundances are found to be in the ranges $0.68\leq A_{\rm O}\leq 0.98$, $0.51\leq A_{\rm Ne}\leq 1.62$, and $0.56\leq A_{\rm Fe}\leq 1.21$, respectively. For Sco~X-1 we freeze the Ne abundance at the solar value because, as previously mentioned, its spectra do not include the Ne-edge absorption region. (Hydrogen column densities will be discussed in Section~\ref{sec_fit_hydrogen}.)

It should be noted that the {\tt TBnew} model only accounts for photoabsorption cross sections of neutral species which, as discussed by \citet{gat13a}, introduces a bias in the broadband-fit parameters such as abundances and H column densities. Regarding pileup effects, the {\tt simple\_gpile2.sl} model gives $\beta$ values in the range $0.041\leq\beta \leq 0.272$ that correspond to pileup degrees in the 11--21~\AA\ region lower than 68\% for Sco~X-1, 31\% for Cygnus~X-1, 17\% for 4U~1820-30,  and 6\% for 4U~1636-53, 4U~1735-44, and GX~9+9. In the oxygen-edge region (21--24~\AA), all sources display pileup degrees lower than 1\% that can be safely ignored although they must still be considered in the broadband-fit stage.

\subsection{Hydrogen column densities}\label{sec_fit_hydrogen}

Hydrogen column densities derived in the broadband fit are compared in Table~\ref{tab3} with previous estimates, in particular two 21~cm surveys and the results by \citet{jue04} who obtained the continuum parameters from fits of the 23.7--24.7~\AA\ line-free region. Good agreement is found except for Cygnus~X-1 and Sco~X-1. In the case of Cygnus~X-1 it is important to note that, as discussed by \citet{han09}, the 21~cm surveys do not resolve $N_{H}$ variations in the region around the source; therefore, we would consider our value to be more reliable. For Sco~X-1 \citet{bra03} found evidence for absorption as a function of X-ray branches (i.e. the position of the source in an X-ray color--color diagram), in a range from $N_{H}=1.1\pm 0.5\times 10^{21}$~cm$^{-2}$ on the horizontal branch to $N_{H}=32\pm 5\times 10^{21}$~cm$^{-2}$ on the normal branch. We conclude that, due to source variability, it is difficult to constrain the hydrogen column density; furthermore, it must also be pointed out that the discrepancies in $N_{H}$ estimates may also arise from the treatment of the source X-ray intrinsic continuum in each model.

\section{Oxygen edge fit}\label{sec_fit_oxygen}

\begin{figure}
  \epsscale{0.9}
  \plotone{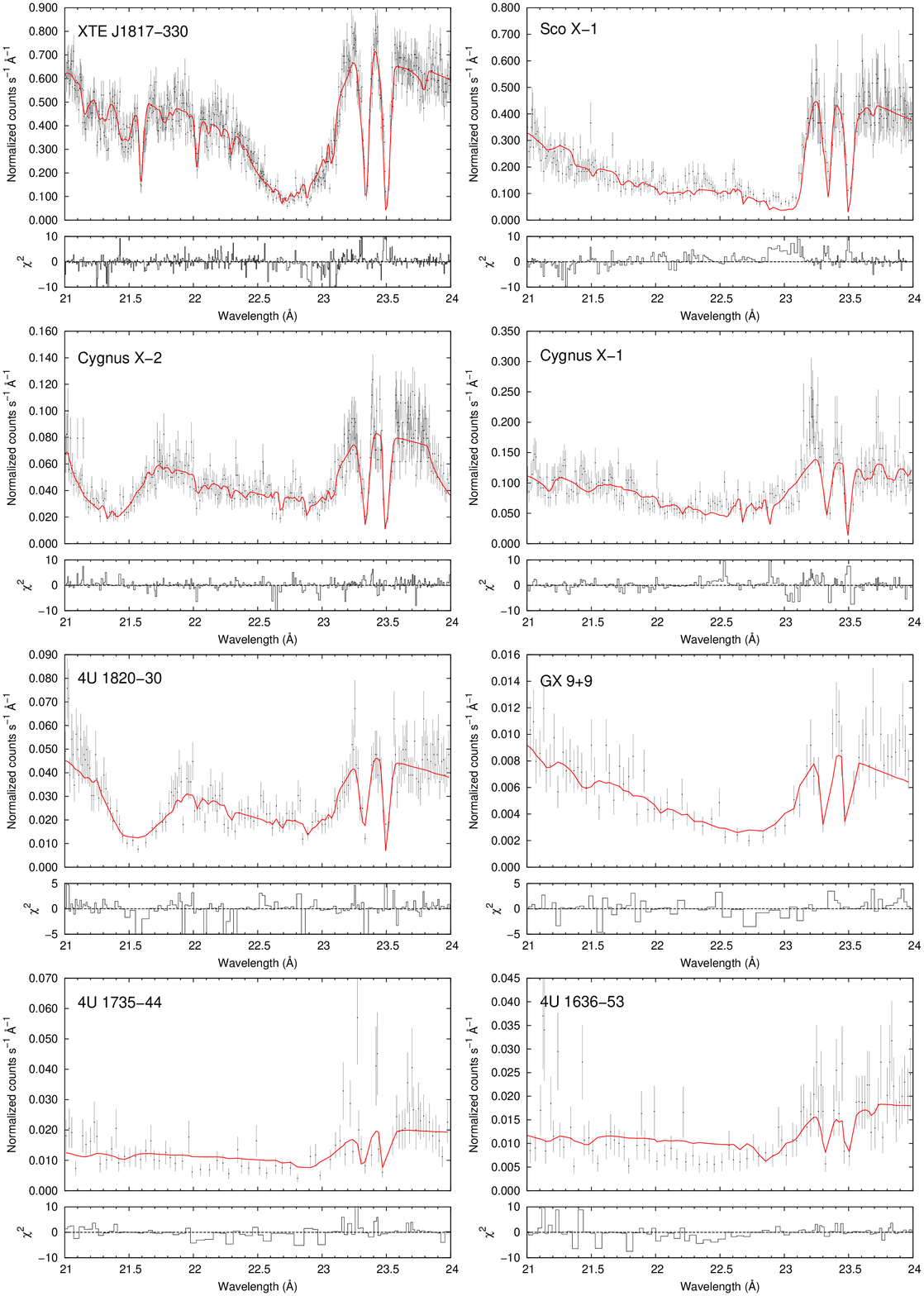}
  \caption{Best fits with the {\tt warmabs} model of the oxygen-edge region in the spectra towards eight Galactic low-mass X-ray binaries. For each source the observation with highest signal-to-noise ratio is shown.\label{fig2}}
\end{figure}

Figure~\ref{fig2} shows the best fits of the oxygen-edge region using the {\tt warmabs}({\tt powerlaw}) model where, for each source, the observation with the highest signal-to-noise is plotted in the 21--24~\AA\ wavelength interval. The dominant features are the \ion{O}{1} and \ion{O}{2} K$\alpha$ absorption lines. Due to the low counts, the detection of absorption lines from resonances with principal quantum number $n>2$ is impaired, the exception being XTE~J1817-330 with good quality spectra that also display K$\alpha$ absorption lines from \ion{O}{6} and \ion{O}{7}. In this respect, \citet{gat13a} have proposed that the \ion{O}{7} K$\alpha$ line in XTE~J1817-330 is formed in the source neighborhood rather than in the ISM. The puzzling absence of the latter line in other sources has been reported by \citet{yao09}, who studied Cygnus~X-2 spectra and suggested the possibility of an \ion{O}{7} resonance emission line annulling its absorption. This effect has been apparently confirmed by \citet{cab13} in analyses of {\it XMM-Newton} RGS and {\it Chandra} MEG spectra, although the exact origin of the emission line is not completely understood.

The best {\tt warmabs} fit parameters using the \ion{O}{1} photoabsorption cross section by \citet{gor13} are given in Table~\ref{tab4}. For all sources the ionization parameter is found to be fairly low, with an average value of $\log\xi=-2.90$ that points to the predominance of neutral and singly ionized oxygen species. There are differences in oxygen abundances among the lines of sight, $0.33\leq A_{\rm O}\leq 1.34$, with an average value of $A_{\rm O}=0.70$ that is consistent with the oxygen abundance reported for XTE~J1817-330 by \citet{gat13a}. If the \ion{O}{1} photoabsorption cross section in {\tt warmabs} is replaced with that by \citet{gar05}, the outcome is not significantly different (see Table~\ref{tab5}): an average ionization parameter of $\log\xi=-2.92$, oxygen abundances in the range of $0.54\leq A_{\rm O}\leq 1.02$, and an abundance average value of $A_{\rm O}=0.68$. It must be clarified here, however, that these O abundances are given relative to the solar standard by \citet{gre98}, but as discussed by \citet{gat13a}, if they are rescaled according to the revision by \citet{asp09}, the abundance average values would be fairly close to solar. The small differences in model output caused by the two photoabsorption cross sections give us confidence that the issue of oxygen atomic data accuracy has been finally settled, and consequently, the data set by \citet{gor13} for \ion{O}{1} can certainly be recommended as the standard in X-ray spectral modeling.

The oxygen abundances for 4U~1735-44 and GX-9+9 by \citet{pin13} when rescaled to the solar standard of \citet{gre98}, respectively $A_{\rm O}=0.67\pm 0.02$ and $A_{\rm O}=0.71\pm 0.02$, are in good agreement with the present values (see Table~\ref{tab4}). On the other hand, for 4U~1820-30 we find $A_{\rm O}=1.34\pm 0.21$, a value greater than that estimated by \citet{cos12} of $A_{\rm O}=0.89\pm 0.02$; however, the hydrogen column densities lead to a similar oxygen column density in both works.

\begin{figure}
  \epsscale{1}
  \plotone{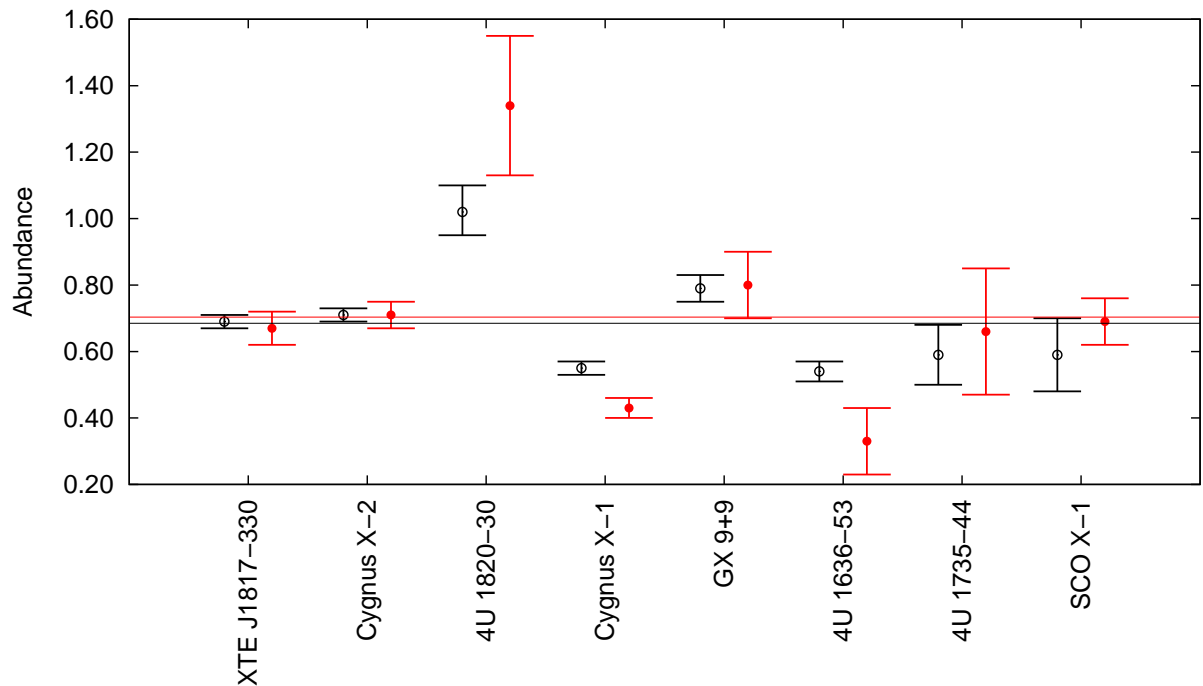}
  \caption{ISM oxygen abundances obtained from the best {\tt warmabs} fits of the spectra towards eight Galactic low-mass X-ray binaries. Open black circles: using the \ion{O}{1} photoabsorption cross section by \citet{gar05}. Filled red circles: using the \ion{O}{1} photoabsorption cross section by \citet{gor13}. Average oxygen abundances of $A_{\rm O}=0.68$ and $A_{\rm O}=0.70$ for these two cases are respectively denoted with black and red horizontal lines. \label{fig3}}
\end{figure}

\begin{figure}
  \epsscale{1}
  \plotone{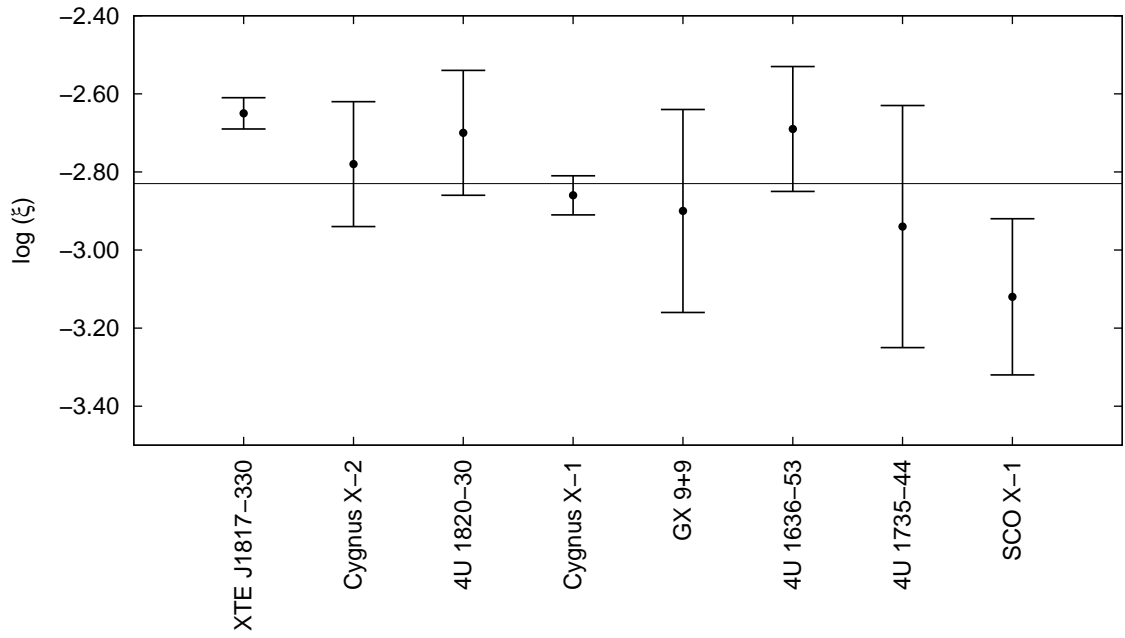}
  \caption{Ionization parameters obtained from the best {\tt warmabs} fits of the spectra towards for eight Galactic low-mass X-ray binaries. The horizontal line corresponds to an average ionization parameter of $\log\xi=-2.908$.\label{fig4}}
\end{figure}

For each source, Figure~\ref{fig3} shows the oxygen abundances obtained with the \ion{O}{1} photoabsorption cross sections by both \citet{gar05} and \citet{gor13}. The sources are ordered according to their total counts (high to low), and the resulting abundance mean values are also shown. Since the abundances obtained for most sources with these two cross sections agree to within the error bars, we proceed our analysis with the most recent \citep{gor13}. The overabundance in the line of sight of 4U~1820-30 is worth noting since it could originate in the source neighborhood in the metal-rich NGC\,6624 globular cluster \citep{guv10}, even though \citet{cos12} have pointed out the difficulty of assigning the precise location of the absorbing gas. In a similar fashion, the ionization parameter for each source is depicted in Figure~\ref{fig4} where a low ionization degree is maintained along the different lines of sight. This result is consistent with the predominant \ion{O}{1} and \ion{O}{2} K$\alpha$ absorption lines observed. Also, the scatter around the mean value is well constrained by the error bars that tend to increase with decreasing count number.

\begin{figure}
  \epsscale{1}
  \plotone{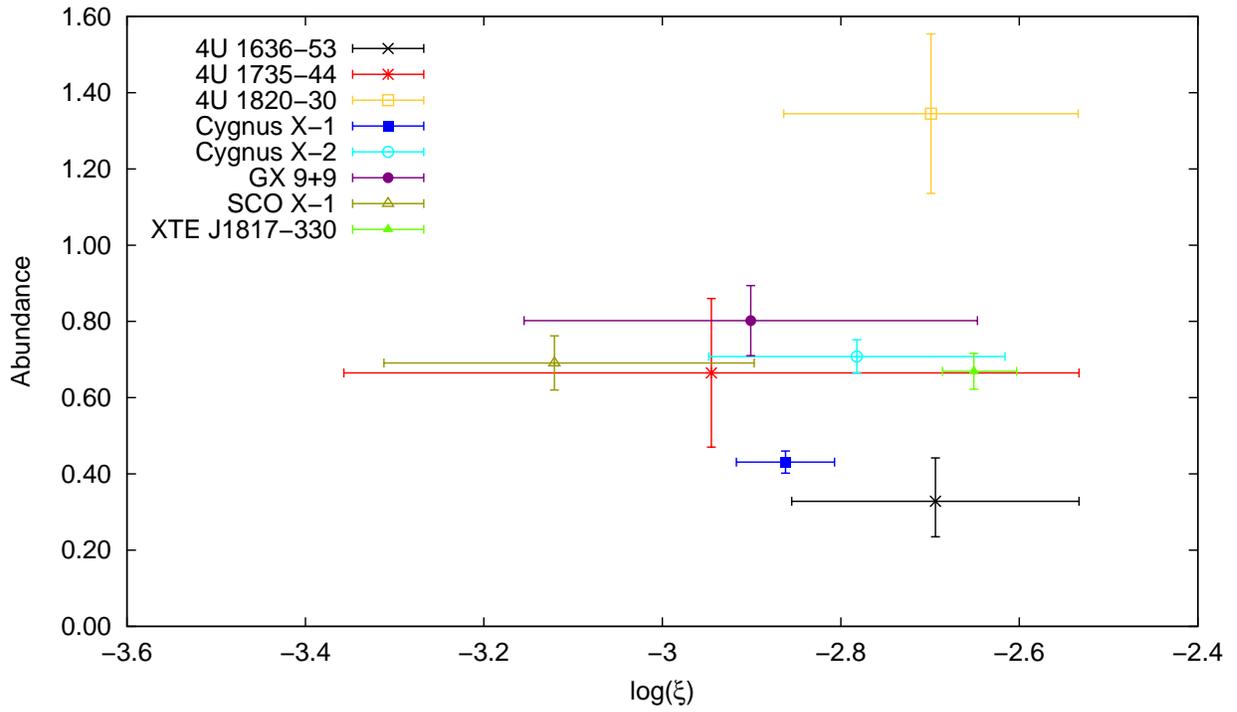}
  \caption{Map of oxygen abundance {\em vs}. ionization parameter denoting the low ionization state of the ISM. Oxygen abundances are mostly lower than solar relative to the standard by \citet{gre98} and close to solar relative to the revision by \citet{asp09}. \label{fig5}}
\end{figure}

\begin{figure}
  \epsscale{1}
  \plotone{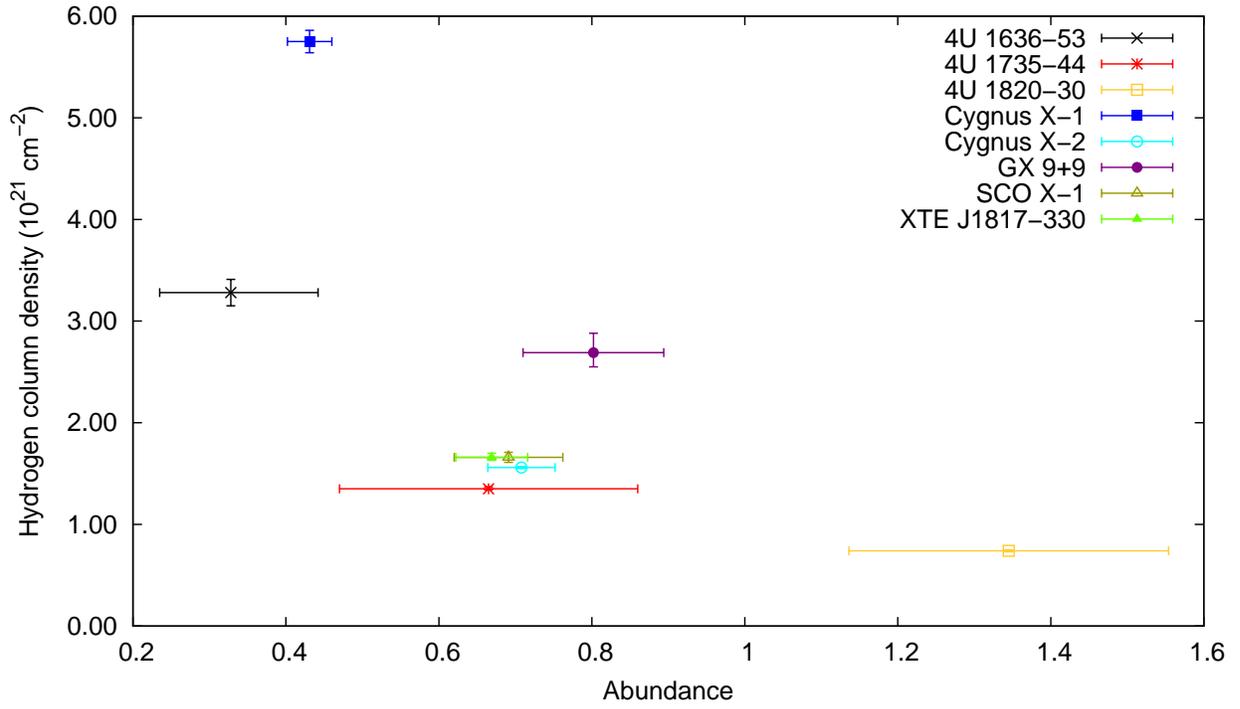}
  \caption{Map of the {\tt TBnew} hydrogen column density {\em vs}. {\tt warmabs} oxygen abundance. The hydrogen column densities are better constrained than the oxygen abundances due to low counts in the oxygen-edge region. \label{fig6}}
\end{figure}

A map of the {\tt warmabs} oxygen abundance {\em vs}. ionization parameter is shown in Figure~\ref{fig5}, where the error bars indicate that the ionization parameter is not as well constrained as the oxygen abundance; this is due to the saturation of the \ion{O}{1} and \ion{O}{2} K$\alpha$ lines. The plot shows that the ISM ionization state is generally low and the oxygen abundance lies below solar except for 4U~1820-30 where its overabundance, as previously mentioned, is likely to be due to its location in the metal-rich NGC\,6624 globular cluster. The map of the {\tt TBnew} hydrogen column density {\em vs}. the {\tt warmabs} oxygen abundance in Figure~\ref{fig6} shows that the hydrogen column density is better constrained, which may a consequence of the low counts in the oxygen-edge region where the {\tt warmabs} fits are performed. The high hydrogen column density of Cygnus~X-1 has been attributed to the presence of an accreted stellar wind along the line of sight \citep{han09}.

\subsection{Oxygen column densities}\label{sec_oxygen_column}

Total oxygen column densities ($N_{\rm O}= A_{\rm O} N_{\rm H}$) are listed in Table \ref{tab6} where the {\tt warmabs} values take into account contributions from \ion{O}{1}, \ion{O}{2}, and \ion{O}{3}. Data by \citet{jue04} are also tabulated, who estimated the oxygen column densities with the optical depth at 21.7 \AA\ and the photoabsorption cross section by \citet{gor00} at that wavelength. Cygnus~X-1 has the highest oxygen column density probably due to the stellar wind from the supergiant companion star as discussed by \citet{han09}. In the case of 4U~1820-30, its low hydrogen column density results in a low oxygen column density despite its high oxygen abundance.

\begin{figure}
  \epsscale{1}
  \plotone{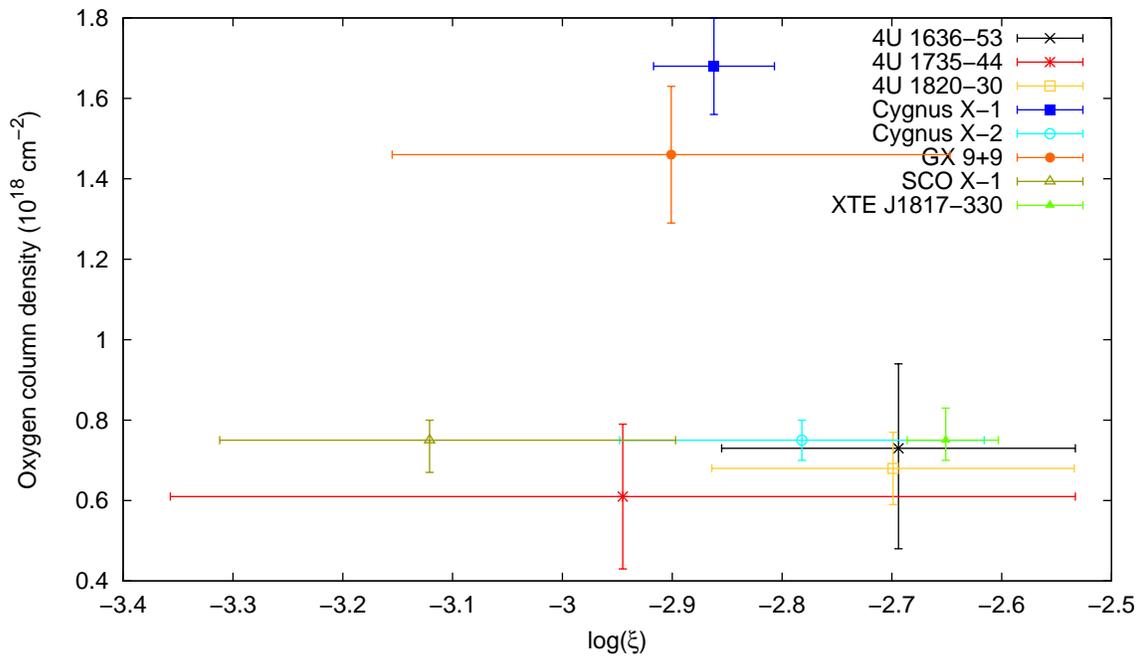}
  \caption{Map of the oxygen column density {\em vs}. ionization parameter generally showing low ionization parameters and low oxygen column densities. \label{fig7}}
\end{figure}

A plot of the total oxygen column density {\em vs.} ionization parameter is depicted in Figure~\ref{fig7}. It shows that the sources appear to have both low ionization parameters and low oxygen column densities; if a correlation does exist between these two properties, a more extensive data analysis would be needed to verify it. For Cygnus~X-1 and GX~9+9, the high hydrogen column densities increase the oxygen column densities although the sources have low ionization parameters. The average oxygen column density of our sampling is $N_{\rm O}=9.2 \times 10^{17}$~cm$^{-2}$.

\section{ISM multiphase structure}\label{sec_ism}

As signatures of a multiphase structure in the ISM, the presence of neutral, medium, and highly ionized atomic species in addition to molecules has been previously studied with high resolution X-ray spectroscopy \citep{yao09, pin10, cos12, pin13, lia13}. They mostly involve the use of Gaussians to fit the observed absorption lines; for instance, \citet{lia13} have recently analyzed 36 {\it Chandra} HETG observations of 11 low-mass X-ray binaries by co-adding all the spectra. For each observation, the central wavelengths of the atomic transitions were corrected for the Galactic rotation velocity relative to the rest frame. Several transitions in the oxygen, magnesium, and neon edges were therein identified, followed by a Bayesian statistical analysis to determine the systematic uncertainties of the measurements. Very high accuracy is reached for the line positions, but in contrast to the present self-consistent model fit, the procedures based on Gaussians are limited regarding line identification and formation. A further disadvantage of such methods when compared to the simultaneous fit lies in the fact that the intrinsic lines with the higher signal-to-noise dominate the co-added spectra. This may lead to the wrong conclusion that each line is present in all observations; as a counterexample, we find that the \ion{O}{6} and \ion{O}{7} K$\alpha$ absorption lines are only present in the XTE~J1817-330 spectra as reported by \citet{gat13a}.

It is known that the use of Gaussian profiles without an underlying analysis of the physical conditions of the gas can lead to incorrect identifications. For instance, Table~4 of \citet{lia13} lists K$\alpha$ absorption lines from \ion{O}{4} and \ion{O}{5} as well as \ion{O}{6} K$\alpha$, K$\beta$, and K$\gamma$; however, as explained by \citet{gat13a}, the neutral component dominates such ISM spectra, and these lines would actually correspond to K$\beta$ and K$\gamma$ transitions in \ion{O}{1} and \ion{O}{2}. In support of this view, present results also point to the dominance of the ISM neutral component in different lines of sight, with a low ionization degree and small variations of the oxygen abundances due to plasma enrichment in the local environment (e.g. the presence of globular clusters). Table~\ref{tab7} lists the \ion{O}{1}, \ion{O}{2}, and \ion{O}{3} column densities for each source as derived from a set of {\sc xstar} simulations using the parameters from the {\tt warmabs} fit. In correspondence with the predicted low ionization parameters, the column density ratios $N$(\ion{O}{2})/$N$(\ion{O}{1}) and $N$(\ion{O}{3})/$N$(\ion{O}{1}) are estimated to be less than 0.1.

Contrary to \citet{lia13}, we only detect \ion{O}{6} K$\alpha$ in XTE~J1817-330. As discussed in \citet{gat13a}, this line probably arises in the source neighborhood, and due to its brightness, an approach using multiple X-ray binaries and spectra addition is likely to assign its origin to the ISM; however, our approach clearly confirms its absence in the rest of the sources. The appearance of highly ionized metal absorption lines in spectra toward Galactic X-ray binaries has been recently considered by \citet{luo14}, who found that most of them arise in the hot gas intrinsic to the source, the ISM only making a small contribution. On the other hand, UV studies have shown the presence of \ion{O}{6} lines towards multiple Galactic lines of sight that would enable us to determine the conditions required to detect \ion{O}{6} absorption in X-ray spectra. For example, \citet{sav03} and \citet{sem03}, using data from the Far-Ultraviolet Spectroscopic Explorer (FUSE), derived \ion{O}{6} average column densities of $8.91\times 10^{13}$~cm$^{-2}$ in the halo region, $2.39\times 10^{14}$~cm$^{-2}$ in the Galactic thick disk, and $1.62\times 10^{14}$~cm$^{-2}$ perpendicular to the Galactic plane. According to the atomic data used in this work, these values would correspond to equivalent widths of $0.288$ m\AA, $0.776$ m\AA, and $0.512$ m\AA, respectively. In this respect, we have performed data simulations using the HETG response files and typical exposure times that have led us to conclude that, for these EW, a statistically acceptable detection would be unsuccessful. In fact, an \ion{O}{6} column density of at least $\approx 1.6\times10^{15}$~cm$^{-2}$ would be required to enable the fit of a Gaussian profile.

An inspection of our fit residuals, by no means negligible, has led to no conclusive evidence regarding photoabsorption by molecules or condensed matter. This may seem polemic as previous studies, even when using comprehensive physical models, have reported that some of the ISM oxygen is indeed depleted to the solid phase. For example, \citet{pin13} have concluded that around 15--25 \% of the oxygen is condensed onto dust. However, it is important to note that in their analysis they include Fe L-edge absorption as well, adding to their models solid oxides such as magnetite (Fe$_{3}$O$_{4}$) and hematite (Fe$_{2}$O$_{3}$). In this respect, our model does lack the two-edge perspective which would perhaps be more reliable in clarifying this contentious issue.

\section{Conclusions}\label{sec_conclusion}

We have carried out an analysis of ISM oxygen K absorption by means of high-resolution X-ray spectra toward eight low-mass Galactic binaries. For this purpose we are employing a physical model, referred to as {\tt warmabs}, that is capable of determining the physical state of the gas. A broadband fit was performed to constrain the hydrogen column density using the {\tt simple\_gpile2.sl}({\tt TBnew}({\tt powerlaw})) model. The pileup effect was estimated from the fit indicating that it must be taken into account in the 11--24~\AA\ region, although it can be safely ignored around the oxygen edge (21--24~\AA). The hydrogen column densities derived from the fits are in good agreement with previous work, although in the case of Cygnus~X-1 it is difficult to constrain it due to the presence of an alleged stellar wind along the line of sight.

For each source, the simultaneous fit of the oxygen edge region with the {\tt warmabs} model provides estimates of the ionization parameter and oxygen abundance. The simultaneous fit is implemented to avoid the overlapping between ISM and intrinsic absorption features thus allowing a reliable study through the line of sight. The low ionization degree is consistent with the prevalence of \ion{O}{1} and \ion{O}{2} K$\alpha$ absorption lines, supporting the dominance of the neutral component in the plasma. The fits yield an average oxygen abundance ($A_{\rm O}=0.70$) lower than solar if the standard by \citet{gre98} is assumed, but fairly close to solar when rescaling to the revision by \citet{asp09}. In the case of 4U~1820-30 oxygen enrichment is encountered, a likely consequence of the source location in the metal-rich NGC\,6624 globular cluster. It is important to note that our conclusions are backed by the present benchmark of oxygen atomic data \citep{gor13} which we now confidently recommend as the standard for future X-ray spectroscopy.

We do not detect high-ionization lines except in XTE~J1817-330 which, as mentioned by \citet{gat13a}, may arise in the source neighborhood rather than in the actual ISM. Using the \ion{O}{6} column densities obtained by \citet{sem03} and \citet{sav03} from UV data, we have carried out a set data simulations to estimate the typical column densities that would enable a statistically acceptable \ion{O}{6} K$\alpha$ line detection, finding that a minimum of $\approx 1.6\times 10^{15}$~cm$^{-2}$ is required to allow the fitting with a Gaussian.

\acknowledgments
Part of this work was carried out by Efra\'in Gatuzz during a visit in January 2014 to the Laboratory of High Energy
Astrophysics, NASA Goddard Space Flight Center, Greenbelt, Maryland, USA. Their hospitality and financial support are  gratefully acknowledged. We would also like to thank Dr. Frits Paerels for comments at the refereeing stage that led to substantial improvements of the arguments presented and thus of the final quality of the paper.



\begin{thebibliography}{47}
\expandafter\ifx\csname natexlab\endcsname\relax\def\natexlab#1{#1}\fi

\bibitem[{{Asplund} {et~al.}(2009){Asplund}, {Grevesse}, {Sauval}, \&
  {Scott}}]{asp09}
{Asplund}, M., {Grevesse}, N., {Sauval}, A.~J., \& {Scott}, P. 2009, \araa, 47,
  481

\bibitem[{{Baumgartner} \& {Mushotzky}(2006)}]{bau06}
{Baumgartner}, W.~H., \& {Mushotzky}, R.~F. 2006, \apj, 639, 929

\bibitem[{{Bautista} \& {Kallman}(2001)}]{bau01}
{Bautista}, M.~A., \& {Kallman}, T.~R. 2001, \apjs, 134, 139

\bibitem[{{Bradshaw} {et~al.}(1999){Bradshaw}, {Fomalont}, \&
  {Geldzahler}}]{Bra99}
{Bradshaw}, C.~F., {Fomalont}, E.~B., \& {Geldzahler}, B.~J. 1999, \apjl, 512,
  L121

\bibitem[{{Bradshaw} {et~al.}(2003){Bradshaw}, {Geldzahler}, \&
  {Fomalont}}]{bra03}
{Bradshaw}, C.~F., {Geldzahler}, B.~J., \& {Fomalont}, E.~B. 2003, \apj, 592,
  486

\bibitem[{{Brandt} {et~al.}(1992){Brandt}, {Castro-Tirado}, {Lund}, {Dremin},
  {Lapshov}, \& {Syunyaev}}]{bra92}
{Brandt}, S., {Castro-Tirado}, A.~J., {Lund}, N., {Dremin}, V., {Lapshov}, I.,
  \& {Syunyaev}, R. 1992, \aap, 262, L15

\bibitem[{{Cabot} {et~al.}(2013){Cabot}, S.~H.~C. and {Wang}, Q.~D. and {Yao}, Y.}]{cab13}
{Cabot}, S.~H.~C. and {Wang}, Q.~D. and {Yao}, Y., R. 2013, \mnras, 431, 511

\bibitem[{{Cackett} {et~al.}(2008){Cackett}, {Miller}, {Raymond}, {Homan}, {van
  der Klis}, {M{\'e}ndez}, {Steeghs}, \& {Wijnands}}]{cac08b}
{Cackett}, E.~M., {Miller}, J.~M., {Raymond}, J., {Homan}, J., {van der Klis},
  M., {M{\'e}ndez}, M., {Steeghs}, D., \& {Wijnands}, R. 2008, \apj, 677, 1233

\bibitem[{{Cackett} {et~al.}(2009){Cackett}, {Miller}, {Homan}, {van der Klis},
  {Lewin}, {M{\'e}ndez}, {Raymond}, {Steeghs}, \& {Wijnands}}]{cac09}
{Cackett}, E.~M., {et~al.} 2009, \apj, 690, 1847

\bibitem[{{Costantini} {et~al.}(2005){Costantini}, {Freyberg}, \&
  {Predehl}}]{cos05}
{Costantini}, E., {Freyberg}, M.~J., \& {Predehl}, P. 2005, \aap, 444, 187

\bibitem[{{Costantini} {et~al.}(2012){Costantini}, {Pinto}, {Kaastra}, {in't
  Zand}, {Freyberg}, {Kuiper}, {M{\'e}ndez}, {de Vries}, \& {Waters}}]{cos12}
{Costantini}, E., {et~al.} 2012, \aap, 539, A32

\bibitem[{{de Vries} \& {Costantini}(2009)}]{dev09}
{de Vries}, C.~P., \& {Costantini}, E. 2009, \aap, 497, 393

\bibitem[{{Dickey} \& {Lockman}(1990)}]{dic90}
{Dickey}, J.~M., \& {Lockman}, F.~J. 1990, \araa, 28, 215

\bibitem[{{Galloway} {et~al.}(2008){Galloway}, {Muno}, {Hartman}, {Psaltis}, \&
  {Chakrabarty}}]{gall08}
{Galloway}, D.~K., {Muno}, M.~P., {Hartman}, J.~M., {Psaltis}, D., \&
  {Chakrabarty}, D. 2008, \apjs, 179, 360

\bibitem[{{Garc{\'{\i}}a} {et~al.}(2005){Garc{\'{\i}}a}, {Mendoza}, {Bautista},
  {Gorczyca}, {Kallman}, \& {Palmeri}}]{gar05}
{Garc{\'{\i}}a}, J., {Mendoza}, C., {Bautista}, M.~A., {Gorczyca}, T.~W.,
  {Kallman}, T.~R., \& {Palmeri}, P. 2005, \apjs, 158, 68

\bibitem[{{Garc{\'{\i}}a} {et~al.}(2011){Garc{\'{\i}}a}, {Ram{\'{\i}}rez},
  {Kallman}, {Witthoeft}, {Bautista}, {Mendoza}, {Palmeri}, \&
  {Quinet}}]{gar11}
{Garc{\'{\i}}a}, J., {Ram{\'{\i}}rez}, J.~M., {Kallman}, T.~R., {Witthoeft},
  M., {Bautista}, M.~A., {Mendoza}, C., {Palmeri}, P., \& {Quinet}, P. 2011,
  \apjl, 731, L15

\bibitem[{{Gatuzz} {et~al.}(2013){Gatuzz}, {Garc{\'{\i}}a}, {Mendoza},
  {Kallman}, {Witthoeft}, {Lohfink}, {Bautista}, {Palmeri}, \&
  {Quinet}}]{gat13a}
{Gatuzz}, E., {et~al.} 2013, \apj, 768, 60

\bibitem[{{Gorczyca} \& {McLaughlin}(2000)}]{gor00}
{Gorczyca}, T.~W., \& {McLaughlin}, B.~M. 2000, J. Phys. B: At. Mol. Opt.
  Phys., 33, L859

\bibitem[{{Gorczyca} {et~al.}(2013){Gorczyca}, {Bautista}, {Hasoglu},
  {Garc{\'{\i}}a}, {Gatuzz}, {Kaastra}, {Kallman}, {Manson}, {Mendoza},
  {Raassen}, {de Vries}, \& {Zatsarinny}}]{gor13}
{Gorczyca}, T.~W., {et~al.} 2013, \apj, 779, 78

\bibitem[{{Grevesse} \& {Sauval}(1998)}]{gre98}
{Grevesse}, N., \& {Sauval}, A.~J. 1998, \ssr, 85, 161

\bibitem[{{G{\"u}ver} {et~al.}(2010){G{\"u}ver}, {Wroblewski}, {Camarota}, \&
  {{\"O}zel}}]{guv10}
{G{\"u}ver}, T., {Wroblewski}, P., {Camarota}, L., \& {{\"O}zel}, F. 2010,
  \apj, 719, 1807

\bibitem[{{Hanke} {et~al.}(2009){Hanke}, {Wilms}, {Nowak}, {Pottschmidt},
  {Schulz}, \& {Lee}}]{han09}
{Hanke}, M., {Wilms}, J., {Nowak}, M.~A., {Pottschmidt}, K., {Schulz}, N.~S.,
  \& {Lee}, J.~C. 2009, \apj, 690, 330

\bibitem[{{Henke} {et~al.}(1993){Henke}, {Gullikson}, \& {Davis}}]{hen93}
{Henke}, B.~L. and {Gullikson}, E.~M. and {Davis}, J.~C. 1993, ADNDT, 54, 181


\bibitem[{{Juett} {et~al.}(2004){Juett}, {Schulz}, \& {Chakrabarty}}]{jue04}
{Juett}, A.~M., {Schulz}, N.~S., \& {Chakrabarty}, D. 2004, \apj, 612, 308

\bibitem[{{Kalberla} {et~al.}(2005){Kalberla}, {Burton}, {Hartmann}, {Arnal},
  {Bajaja}, {Morras}, \& {P{\"o}ppel}}]{kal05}
{Kalberla}, P.~M.~W., {Burton}, W.~B., {Hartmann}, D., {Arnal}, E.~M.,
  {Bajaja}, E., {Morras}, R., \& {P{\"o}ppel}, W.~G.~L. 2005, \aap, 440, 775

\bibitem[{{Liao} {et~al.}(2013){Liao}, {Zhang}, \& {Yao}}]{lia13}
{Liao}, J.-Y., {Zhang}, S.-N., \& {Yao}, Y. 2013, \apj, 774, 116

\bibitem[{{Lodders,} {K}(2003){Lodders}}]{lod03}
{Lodders}, K. 2003, \apj, 591, 1220

\bibitem[{{Luo} \& {Fang}(2014)}]{luo14}
{Luo}, Y., \& {Fang}, T. 2014, \apj, 780, 170

\bibitem[{{Makishima} {et~al.}(2008){Makishima}, {Takahashi}, {Yamada}, {Done},
  {Kubota}, {Dotani}, {Ebisawa}, {Itoh}, {Kitamoto}, {Negoro}, {Ueda}, \&
  {Yamaoka}}]{mak08}
{Makishima}, K., {et~al.} 2008, \pasj, 60, 585

\bibitem[{{Massey} {et~al.}(1995){Massey}, {Johnson}, \&
  {Degioia-Eastwood}}]{mas95}
{Massey}, P., {Johnson}, K.~E., \& {Degioia-Eastwood}, K. 1995, \apj, 454, 151

\bibitem[{{McLaughlin} {et~al.}(2013){McLaughlin}, {Ballance}, {Bowen}, {Gardenghi},\&
  {Stolte}}]{mcl13}
{McLaughlin}, B.~M. and {Ballance}, C.~P. and {Bowen}, K.~P. and
	{Gardenghi}, D.~J. and {Stolte}, W.~C. 2013, \apjl, 771, L8


\bibitem[{{Migliari} {et~al.}(2004){Migliari}, {Fender}, {Rupen}, {Wachter},
  {Jonker}, {Homan}, \& {van der Klis}}]{mig04}
{Migliari}, S., {Fender}, R.~P., {Rupen}, M., {Wachter}, S., {Jonker}, P.~G.,
  {Homan}, J., \& {van der Klis}, M. 2004, \mnras, 351, 186

\bibitem[{{Miller} {et~al.}(2009){Miller}, {Cackett}, \& {Reis}}]{mil09b}
{Miller}, J.~M., {Cackett}, E.~M., \& {Reis}, R.~C. 2009, \apjl, 707, L77

\bibitem[{{Miller} {et~al.}(2006){Miller}, {Raymond}, {Homan}, {Fabian},
  {Steeghs}, {Wijnands}, {Rupen}, {Charles}, {van der Klis}, \&
  {Lewin}}]{mil06a}
{Miller}, J.~M., {et~al.} 2006, \apj, 646, 394

\bibitem[{{Nousek} \& {Shue}(1989)}]{nou89}
{Nousek}, J.~A., \& {Shue}, D.~R. 1989, \apj, 342, 1207

\bibitem[{{Orosz} \& {Kuulkers}(1999)}]{oro99}
{Orosz}, J.~A., \& {Kuulkers}, E. 1999, \mnras, 305, 132

\bibitem[{{Pandel} {et~al.}(2008){Pandel}, {Kaaret}, \& {Corbel}}]{pan08}
{Pandel}, D., {Kaaret}, P., \& {Corbel}, S. 2008, \apj, 688, 1288

\bibitem[{{Pinto} {et~al.}(2013){Pinto}, {Kaastra}, {Costantini}, \& {de
  Vries}}]{pin13}
{Pinto}, C., {Kaastra}, J.~S., {Costantini}, E., \& {de Vries}, C. 2013, \aap,
  551, A25

\bibitem[{{Pinto} {et~al.}(2010){Pinto}, {Kaastra}, {Costantini}, \&
  {Verbunt}}]{pin10}
{Pinto}, C., {Kaastra}, J.~S., {Costantini}, E., \& {Verbunt}, F. 2010, \aap,
  521, A79

\bibitem[{{Sala} \& {Greiner}(2006)}]{sal06}
{Sala}, G., \& {Greiner}, J. 2006, The Astronomer's Telegram, 791, 1

\bibitem[{{Savage} {et~al.}(2003){Savage}, {Sembach}, {Wakker}, {Richter},
  {Meade}, {Jenkins}, {Shull}, {Moos}, \& {Sonneborn}}]{sav03}
{Savage}, B.~D., {et~al.} 2003, \apjs, 146, 125

\bibitem[{{Schattenburg} \& {Canizares}(1986)}]{sch86}
{Schattenburg}, M.~L., \& {Canizares}, C.~R. 1986, \apj, 301, 759

\bibitem[{{Schulz} {et~al.}(2002){Schulz}, {Cui}, {Canizares}, {Marshall},
  {Lee}, {Miller}, \& {Lewin}}]{sch02}
{Schulz}, N.~S., {Cui}, W., {Canizares}, C.~R., {Marshall}, H.~L., {Lee},
  J.~C., {Miller}, J.~M., \& {Lewin}, W.~H.~G. 2002, \apj, 565, 1141

\bibitem[{{Sembach} {et~al.}(2003){Sembach}, {Wakker}, {Savage}, {Richter},
  {Meade}, {Shull}, {Jenkins}, {Sonneborn}, \& {Moos}}]{sem03}
{Sembach}, K.~R., {et~al.} 2003, \apjs, 146, 165


\bibitem[{{Stolte} {et~al.}(1997){Stolte}, {Lu}, {Samson}, {Hemmers},
  {Hansen}, {Whitfield}, {Wang}, {Glans}, \& {Lindle}}]{sto97}
{Stolte}, K.~R., {et~al.} 1997, JPhB, 30, 4489



\bibitem[{{Takei} {et~al.}(2002){Takei}, {Fujimoto}, {Mitsuda}, \&
  {Onaka}}]{tak02}
{Takei}, Y., {Fujimoto}, R., {Mitsuda}, K., \& {Onaka}, T. 2002, \apj, 581, 307

\bibitem[{{Verner} {et~al.}(1996){Verner}, {Ferland}, {Korista}, \&
  {Yakovlev}}]{ver96}
{Verner}, D.~A., {Ferland}, G.~J., {Korista}, K.~T., \& {Yakovlev}, D.~G. 1996,
  \apj, 465, 487

\bibitem[{{Wilms} {et~al.}(2000){Wilms}, {Allen}, \& {McCray}}]{wil00}
{Wilms}, J., {Allen}, A., \& {McCray}, R. 2000, \apj, 542, 914

\bibitem[{{Yao} {et~al.}(2009){Yao}, {Schulz}, {Gu}, {Nowak}, \&
  {Canizares}}]{yao09}
{Yao}, Y., {Schulz}, N.~S., {Gu}, M.~F., {Nowak}, M.~A., \& {Canizares}, C.~R.
  2009, \apj, 696, 1418

\bibitem[{{Yao} \& {Wang}(2005)}]{yao05}
{Yao}, Y., \& {Wang}, Q.~D. 2005, \apj, 624, 751

\bibitem[{{Yao} \& {Wang}(2006)}]{yao06}
---. 2006, \apj, 641, 930

\bibitem[{{Zdziarski} {et~al.}(2004){Zdziarski}, {Gierli{\'n}ski},
  {Miko{\l}ajewska}, {Wardzi{\'n}ski}, {Smith}, {Harmon}, \&
  {Kitamoto}}]{zdz04}
{Zdziarski}, A.~A., {Gierli{\'n}ski}, M., {Miko{\l}ajewska}, J.,
  {Wardzi{\'n}ski}, G., {Smith}, D.~M., {Harmon}, B.~A., \& {Kitamoto}, S.
  2004, \mnras, 351, 791

\end{thebibliography}


\clearpage
\newpage

\clearpage
\newpage

\clearpage
\newpage

\clearpage
\newpage

\begin{deluxetable}{lllcllll}
\tabletypesize{\scriptsize}
\tablecaption{{\em Chandra} observations used in the present work \label{tab1}}
\tablewidth{0pt}
\tablehead{
\colhead{Source}  &\colhead{ObsID} & \colhead{Date} & \colhead{Exposure} & \colhead{Read Mode} &\colhead{Galactic}&\colhead{Distance}\\
\colhead{ }  &\colhead{ } &\colhead{ } &\colhead{(ks)}&\colhead{ } &\colhead{Coordinates}&\colhead{ }&\colhead{ }

}
\startdata
 4U~1636-53&105&1999 Oct 20 & 29 &TIMED & $(332.9,-4.8)$  & 5.95~kpc$^{a}$\\
& 1939&2001 Mar 28 & 27 &TIMED & \\
&6635&2006 Mar 22 & 23 &	CONTINUOUS &   \\
&6636&2007 Jul 02 & 26 &CONTINUOUS  & \\
4U~1735-44&704&2000 Jun 09 & 24  &TIMED  & $(346.0,-6.9)$ & 6.5~kpc$^{b}$\\
&6637&2006 Aug 17 & 25 &	CONTINUOUS  & \\
&6638&2007 Mar 15 & 23 &CONTINUOUS  & \\
4U~1820-30&1021&2001 Jul 21 & 9.6 &TIMED  & $(2.7,-7.9)$ & 7.6~kpc$^{c}$\\
&1022&2001 Sep 12 & 11 &	TIMED   & \\
&6633&2006 Aug 12 & 25 &CONTINUOUS  & \\
&6634&2006 Oct 20 & 26 &CONTINUOUS  & \\
&7032&2006 Nov 05 & 47 &CONTINUOUS  & \\
Cygnus~X-1&3407&2001 Oct 28 & 17 &	CONTINUOUS  & $(71.3, 3.0)$ & 7.2~kpc$^{d}$ \\
&3724&2002 Jul 30 & 8.8 &CONTINUOUS  & \\
Cygnus~X-2&8170&2007 Aug 25 & 65 &	CONTINUOUS & $(87.3,-11.3)$ & 2.14~kpc$^{e}$\\
&8599&2007 Aug 23 & 59 &CONTINUOUS & \\
&10881&2009 May 12 & 66 &CONTINUOUS  & \\
&1102&1999 Sep 23 & 28  &TIMED  &\\
GX~9+9&703&2000 Aug 22 & 20 &TIMED& $(8.5, 9.0)$  & 8~kpc$^{f}$ \\
&11072&2010 Jul 13 & 95 &TIMED& \\
Sco~X-1&3505&2003 Jul 21    &  16  &CONTINUOUS& $(359, 23.7)$    & 2.8~kpc$^{g}$ \\
XTE~J1817-330&6615&2006 Feb 13&18 &CONTINUOUS& 	$(359.8, -7.9)$ & 1--4~kpc$^{h}$ \\
&6616&2006 Feb 24&29 &CONTINUOUS& \\
&6617&2006 Mar 15&47&CONTINUOUS& \\
&6618&2006 May 22&51&CONTINUOUS& \\
\enddata
\tablecomments{Distances are taken from $^a$\citet{bra92}; $^b$\citet{gall08}; $^c$\citet{mig04}; $^d$\citet{oro99}; $^e$\citet{mas95}; $^f$\citet{zdz04}; $^g$\citet{Bra99}; and $^h$\citet{sal06}.}
\end{deluxetable}


\clearpage
 \begin{deluxetable}{lllllcc}
\tabletypesize{\scriptsize}
 \tablecaption{Broadband simultaneous fit parameters \label{tab2}}
\tablewidth{0pt}
\tablehead{
\colhead{Source} &\multicolumn{4}{c}{\tt TBnew}  &\colhead{{\tt SimpleGpile2}$^{a}$} & \colhead{Reduced chi-square }\\
\cline{2-5}\\
\colhead{} & \colhead{$N_{\rm H}$ ($10^{21}$~cm$^{-2}$)} & \colhead{A$_{\rm Ne}$}& \colhead{A$_{\rm Fe}$}& \colhead{A$_{\rm O}$}& \colhead{$\beta$ }& \colhead{$\chi^{2}$ }
}
\startdata
4U~1636-53 &$3.08\pm 0.23$ &$1.54\pm 0.17$ & $1.09\pm 0.12$ & $0.79\pm 0.09$&$ 0.049 \pm  0.005$&$1.22$\\
4U~1735~44 &$1.35\pm 0.02$&$1.02\pm 0.12$ & $0.96^{ + 0.08} _{ - 0.12}$&$0.98\pm 0.02$&$ 0.041 \pm  0.004$&$1.10$\\
4U~1820~30 &$0.74\pm 0.01$&$0.77 ^{+ 0.18 } _{- 0.07}$ & $1.21\pm 0.09$&$0.68\pm 0.02$&$ 0.050 \pm 0.005$&$1.12$\\
Cygnus~X-1 &$ 5.75\pm 0.11$&$1.54\pm 0.05$ &$ 0.56\pm 0.02$ &$0.68 \pm 0.02$&$ 0.049 \pm 0.005$ &$1.22$\\
Cygnus~X-2 &$1.56\pm 0.01$&$0.51 \pm 0.02$ & $1.07\pm 0.02$&$0.71 \pm 0.01$&$ 0.048 \pm 0.005$&$1.41$\\
GX~9+9 &$2.69_{-0.14}^{+0.19}$&$0.83 \pm 0.14$ &$ 1.05 \pm 0.11$ &$0.73 \pm 0.05$&$ 0.042  \pm 0.004 $&$ 1.10$\\
Sco~X-1&$1.66\pm 0.06 $&$1.0$ (fixed) & $0.63\pm 0.13$ &$0.76 \pm 0.07$&$0.272\pm 0.005   $&$1.22  $\\
XTE~J1817-330$^{b}$ &$1.66_{-0.04}^{+0.03}$&$ 1.62\pm 0.05$ & $1.10\pm 0.04$&$0.84 ^{+ 0.014} _{ - 0.01}$&$0.050\pm 0.001$&$1.20$\\
\enddata
\tablecomments{Abundances relative to the solar values of \citet{gre98}.}
\tablenotetext{a}{Mean value}
\tablenotetext{b}{Data from \citet{gat13a}}
\end{deluxetable}


\clearpage
\begin{deluxetable}{llcclll}
\tabletypesize{\scriptsize}
 \tablecaption{Hydrogen column densities \label{tab3}}
\tablewidth{0pt}
\tablehead{
 \colhead{Source}&  \colhead{Present Work$^{a}$} & \colhead{$21$~cm} & \colhead{$21$~cm}  & \colhead{JSC$^{d}$} & \colhead{Others}\\
 \colhead{} &\colhead{}&\colhead{Survey$^b$}&\colhead{Survey$^c$}&\colhead{}  &\colhead{} &\colhead{}
}
\startdata
4U~1636-53  &$3.08\pm 0.23$ & 3.30 & 2.64 & $5.3^{+2.1}_{-0.7}$ & $3.9\pm1.2^{e}$\\
4U~1735~44  & $1.35\pm 0.02$  & 3.03& 2.56& $7\pm 2$&$3.0^{e}$\\
4U~1820~30  & $0.74\pm 0.01 $ &1.52&1.32&$2.7 ^{+0.4}_{-0.3}$&$0.78\pm 0.03^{f}$\\
Cygnus~X-1 &$5.75\pm 0.11  $&8.10&7.81& $5.35\pm 0.6$ &  $6.6 ^{+0.8}_{-0.3}$ $^{g}$\\
&&&&& $5.4\pm 0.4^{h}$\\
Cygnus~X-2  & $1.56\pm 0.01$ &2.20 & 1.88 &$2.3\pm 0.5$  &       $1.9\pm0.5^{i}$            \\
GX~9+9  & $ 2.69_{-0.14}^{+0.19}$ &2.10 &1.98  & $4.5^{+2.4}_{-1.2}$  &                 \\
Sco~X-1  &$1.66\pm 0.06 $ &1.47 &1.40  &    &  $1.33\pm 0.02$    $^{j}$           \\
&&&&&$1.1\pm 0.5$ (HB)$^{k}$\\
&&&&& $ 32\pm 5$ (NB)$^{k}$\\
XTE~J1817-330  &$1.66_{-0.04}^{+0.03}$ $^{l}$ &1.58 &1.39     &               \\
\enddata
\tablecomments{$N_{\rm H}$ in units of $10^{21}$~cm$^{-2}$.}
\tablenotetext{a}{{\tt TBnew} fit}
\tablenotetext{b}{\citet{dic90}}
\tablenotetext{c}{\citet{kal05}}
\tablenotetext{d}{\citet{jue04}; for Cygnus~X-1, the average is given}
\tablenotetext{e}{\citet{cac09}}
\tablenotetext{f}{\citet{cos12}}
\tablenotetext{g}{\citet{mak08}}
\tablenotetext{h}{\citet{han09}}
\tablenotetext{i}{\citet{cos05}}
\tablenotetext{j}{\citet{gar11}}
\tablenotetext{k}{\citet{bra03}; HB = horizontal branch, NB = normal branch}
\tablenotetext{l}{\citet{gat13a}}
\end{deluxetable}


\begin{deluxetable}{llll}
 \tabletypesize{\scriptsize}
 \tablecaption{{\tt warmabs} fit parameters \label{tab4}}
\tablewidth{0pt}
\tablehead{
\colhead{Source}  & \multicolumn{2}{c}{\tt warmabs}   &\colhead{Statistics}   \\
\colhead{ }  & \colhead{log $\xi$} & \colhead{A$_{\rm O}$}  &\colhead{$\chi^2/{\rm dof}$}
}
\startdata
4U~1636-53    &$-2.69\pm 0.16$ &$0.33\pm 0.10$ &$167/102$  \\
4U~1735-44    &$-2.94\pm 0.31$ &$0.66\pm 0.19$ &$126/101$   \\
4U~1820-30    &$-2.70\pm 0.16$ &$1.34\pm 0.21$ &$ 479/436$  \\
Cygnus~X-1    &$-2.86\pm 0.05$ &$0.43\pm 0.03$ &$594/442 $  \\
Cygnus~X-2    &$-2.78\pm 0.16$ &$0.71\pm 0.04$ &$1208/1058$  \\
GX~9+9        &$-2.90\pm 0.26$ &$0.80\pm 0.10$ &$123/130$    \\
Sco~X-1       &$-3.12\pm 0.20$ &$0.69\pm 0.07$ &$ 326/273 $  \\
XTE~J1817-330 &$-2.65\pm 0.04$ &$0.67\pm 0.05$ &$2309/1865$  \\
\enddata
\tablecomments{Fits carried out with the O~{\sc i} photoabsorption cross section by \citet{gor13}. Oxygen abundances relative to the solar value of \citet{gre98}.}
\end{deluxetable}


\begin{deluxetable}{llll}
 \tabletypesize{\scriptsize}
 \tablecaption{{\tt warmabs} fit parameters \label{tab5}}
\tablewidth{0pt}
\tablehead{
\colhead{Source}  & \multicolumn{2}{c}{\tt warmabs} & \colhead{Statistics}  \\
\colhead{ }  & \colhead{log $\xi$} & \colhead{A$_{\rm O}$} & \colhead{$\chi^2/{\rm dof}$}
}
\startdata
4U~1636-53         &$-3.29^{+0.10}_{-0.23}$  &$0.54\pm 0.03$    &$164/102$  \\
4U~1735-44         &$-2.67^{+0.14}_{-0.52}$  &$0.59\pm 0.09$    &$126/101$  \\
4U~1820-30         &$-2.68^{+0.14}_{-0.01}$  &$1.02\pm 0.08$    &$ 486/436$ \\
Cygnus~X-1         &$-3.35\pm 0.08 $         &$0.55\pm 0.02$    &$619/442 $  \\
Cygnus~X-2         &$-2.76\pm 0.06$          &$0.71\pm 0.02$    &$1228/1058$ \\
GX~9+9             &$-2.86\pm 0.16$          &$0.79\pm 0.04$    &$124/130$   \\
Sco~X-1            &$-3.10^{+0.24}_{-0.17}$  &$0.59\pm 0.11$    &$ 330/273 $ \\
XTE~J1817-330$^{a}$&$-2.69\pm 0.02$          &$0.68\pm 0.01$    &$2321/1865$ \\
\enddata
\tablecomments{Fits carried out with the O~{\sc i} photoabsorption cross section by \citet{gar05}. Oxygen abundances relative to the solar value of \citet{gre98}.}
\tablenotetext{a}{Data from \citet{gat13a}}
\end{deluxetable}


\clearpage
\begin{deluxetable}{llll}
\tabletypesize{\scriptsize}
 \tablecaption{Total oxygen column densities \label{tab6}}
\tablewidth{0pt}
\tablehead{
 \colhead{Source}& \colhead{{\tt warmabs}$^{a}$} & \colhead{JSC$^{b}$ }   & \colhead{Others}
}
\startdata
4U~1636-53    &$0.73^{+0.20}_{-0.26}$  &$2.6^{+1.0}_{-0.3}$     &$2.40\pm 0.05^c$          \\
4U~1735~44    &$0.61\pm 0.17$          &$3.4\pm 1.2$            &                          \\
4U~1820~30    &$0.68\pm 0.03$          &$1.31 ^{+0.20}_{-0.14}$ &$0.31^{+0.30}_{-0.15}$ $^d$  \\
              &                        &                        &$0.98\pm 0.10^e$          \\
Cygnus~X-1    & $1.68^{+0.04}_{-0.02}$ &$2.65 \pm 0.30$         &$3.92 \pm 0.23^f$         \\
              &                        &                        &$3.81^{+0.04}_{-0.03}$ $^g$  \\
Cygnus~X-2    &$0.75^{+0.02}_{-0.03}$  &$1.1\pm 0.2$            &$0.86\pm 0.28^h$          \\
              &                        &                        &$0.91^{+0.15}_{-0.18}$ $^i$  \\
GX~9+9        &$1.46^{+0.17}_{-0.18}$  &$ 2.2 ^{+1.2}_{-0.6}$   &                          \\
Sco~X-1       &$0.75^{+0.05}_{-0.08}$  &                        &$0.90 \pm 0.12^j$         \\
XTE~J1817-330 &$0.75\pm 0.10$          &                        &                          \\
\enddata
\tablecomments{$N_{\rm O}$ in units of $10^{18}$~cm$^{-2}$.}
\tablenotetext{a}{Derived from the {\tt warmabs} fit (see Section~\ref{sec_oxygen_column})}
\tablenotetext{b}{\citet{jue04}; for Cygnus~X-1, the average is given}
\tablenotetext{c}{\citet{pan08}, average value}
\tablenotetext{d}{\citet{yao06}, warm phase}
\tablenotetext{e}{\citet{cos12}}
\tablenotetext{f}{\citet{sch02}}
\tablenotetext{g}{\citet{han09}}
\tablenotetext{h}{\citet{tak02}}
\tablenotetext{i}{\citet{yao09}}
\tablenotetext{j}{\citet{gar11}}
\end{deluxetable}


\clearpage
\begin{deluxetable}{llll}
\tabletypesize{\scriptsize}
 \tablecaption{Partial oxygen column densities \label{tab7}}
\tablewidth{0pt}
\tablehead{
 \colhead{Source}& \colhead{O~{\sc i}}& \colhead{O~{\sc ii}}& \colhead{O~{\sc iii}}
}
\startdata
4U~1636-53    &$7.01\pm	2.13$&$	0.30\pm	0.09$&$0.0070\pm 	0.0022$         \\
4U~1735~44    &$5.92\pm 1.71$&$0.10\pm0.03$&	$0.0010\pm 	0.0003$ \\
4U~1820~30    &$6.42\pm	1.01$&$	0.27\pm0.04$	&$0.0060\pm	0.0010$\\
Cygnus~X-1    &$16.34\pm 1.14$&$0.36\pm0.02$&$	0.0050\pm	0.0004$\\
Cygnus~X-2    &$7.26\pm 0.40$&$0.22\pm0.01$&$	0.0040\pm	0.0002$\\
GX~9+9        &$14.27\pm 1.78$&$0.27\pm0.03$&$	0.0030\pm	0.0004$\\
Sco~X-1       &$7.67\pm	0.77$&$	0.06\pm	0.01$&$	0.0010\pm	0.0001$\\
XTE~J1817-330 &$7.14\pm	0.53$&$	0.36\pm	0.03$&$	0.0100\pm	0.0008$\\
\enddata
\tablecomments{$N_{\rm O}$ in units of $10^{17}$~cm$^{-2}$.}
\end{deluxetable}

\end{document}